\documentclass[12pt,preprint]{aastex}

\shorttitle{PROPER MOTION OF THE MAGNETAR 1E~1547.0--5408}
\shortauthors{DELLER ET AL.}

\begin{document}

\def\mag{PSR~J1550--5418}
\def\cal{J1515--5559}

\title{The proper motion of PSR~J1550--5418 measured with VLBI: a second magnetar velocity measurement}

\author{A.~T.~Deller\altaffilmark{1}, F.~Camilo\altaffilmark{2},
J.~E.~Reynolds\altaffilmark{3}, and J.~P.~Halpern\altaffilmark{2}}

\altaffiltext{1}{ASTRON, P.O. Box 2, 7990 AA Dwingeloo, The Netherlands}

\altaffiltext{2}{Columbia Astrophysics Laboratory, Columbia University,
  550 West 120th Street, New York, NY~10027, USA}

\altaffiltext{3}{Australia Telescope National Facility, CSIRO, Epping,
  NSW~1710, Australia}

\begin{abstract}

The formation mechanism of neutron stars with extremely large magnetic
field strengths (magnetars) remains unclear.  Some formation scenarios
predict that magnetars should be born with extremely high space velocities,
$>1000$\,km\,s$^{-1}$.  Using the Long Baseline Array in Australia,
we have measured the proper motion of the intermittently radio-bright
magnetar \mag\ (1E~1547.0--5408): $\mu = 9.2\pm0.6$\,mas\,yr$^{-1}$.
For a likely distance of $6\pm2$\,kpc, the implied transverse velocity
is $280^{+130}_{-120}$\,km\,s$^{-1}$ after correcting for Galactic rotation.  
Along with the $\approx 200$\,km\,s$^{-1}$
transverse velocity measured for the magnetar XTE~J1810--197, this result
suggests that formation pathways producing large magnetic fields
do not require very large birth kicks.

\end{abstract}


\section{Introduction}
\label{sec:intro}

Magnetars, originally identified via their bright and variable
high-energy emission, are neutron stars largely powered by
the decay of ultra-strong magnetic fields \citep{dt92a}.  Known examples of
magnetars\footnote{http://www.physics.mcgill.ca/$\sim$pulsar/magnetar/main.html.}
-- both anomalous X-ray pulsars (AXPs) and soft gamma-ray repeaters (SGRs) --
are young, but rotate slowly because their strong dipole fields
($B_s \sim 10^{14-15}$\,G) have spun them down quickly.
\citet{dt92a} and \citet{td93a}
suggested that such large fields can only be created through
dynamo action in proto-neutron stars with $\sim 1$\,ms spin periods,
which is shorter than the convective overturn time.
Searches for evidence of such fast initial rotation via its
enhanced contribution, up to $10^{52}$ erg,
to the explosion energy of the supernova 
remnants (SNRs) hosting magnetars, have generally come up negative
\citep{vin06,hal10}.   On the other hand, it is possible that
the diffusion of such rotational energy through the envelope
of a supernova can explain some of the most luminous
SNe light curves that have been observed in recent years \citep{kas10}.
The birth of a millisecond magnetar is also considered
a possible channel for gamma-ray bursts.

A possible consequence of the large internal magnetic fields ($B \ge 10^{15}$\,G)
or fast initial rotation ($P \le 1$\,ms) of magnetars is a very high kick velocity.
Processes that are ineffective in ordinary neutron stars, but
may produce $v>1000$\,km\,s$^{-1}$ recoils in magnetars, include
anisotropic mass loss in a wind or jet, the electromagnetic rocket effect,
and anisotropic neutrino emission \citep{dt92a,td93a}.  On the other hand,
the central location of some magnetars within SNRs
has led to the inference that as a class their velocity is $<500$\,km\,s$^{-1}$
\citep{gsgv01}, closer to that of the population of ordinary pulsars.
The superposition of SGRs on massive star clusters
or giant molecular clouds \citep{cor04} that are considered 
their likely birth places also offers circumstantial evidence
that their velocities are unremarkable.

Attempts to actually measure velocities via X-ray or infrared imaging remain
unfulfilled, and only
the first radio-detected magnetar, XTE~J1810--197, has a reported
proper motion.  Using VLBA observations, \citet{helfand07a} obtained a
transverse velocity of $\approx 200$\,km\,s$^{-1}$ for XTE~J1810--197,
slightly below the average for young neutron stars.  Although the unknown
radial component of velocity makes it impossible to draw firm conclusions,
this result is not supportive of theories of magnetar formation mechanisms that 
require a high initial velocity. However, more measurements are required in order to
actually measure the initial velocity distribution of this exotic class of neutron stars.

The discovery of a second radio-emitting magnetar, \mag\ \citep{crhr07},
provided the opportunity for a second proper motion measurement.
Here we present Long Baseline Array observations and the resulting proper
motion for this 2\,s pulsar (also known as 1E~1547.0--5408).  Our VLBI
observations and results are presented in Sections~\ref{sec:vlbi} and
\ref{sec:results}, and the implications for the birth velocity of \mag\
and magnetars in general are discussed in Section~\ref{sec:velocity}.

\section{VLBI observations}
\label{sec:vlbi}

We observed \mag\ five times over a period of 2.5 years using Australia's
Long Baseline Array (LBA).  The LBA consists of six antennas: the Australia
Telescope Compact Array (ATCA), Parkes, Mopra and Tidbinbilla antennas
operated by CSIRO Astronomy and Space Sciences, and the Hobart and Ceduna
antennas operated by the University of Tasmania.  The first observation,
obtained using a target of opportunity trigger in 2007 August, was made at
2.3\,GHz, and revealed that \mag\ suffers from severe scatter broadening
due to the intervening ionized interstellar medium \citep[already measured
in the time domain by][]{crhr07}.  \mag\ was detected on the shortest baselines
only, and the scatter broadening was estimated to be approximately 80
milliarcseconds (mas) at 2.3\,GHz.  This scatter-broadening exceeds the
prediction of the NE2001 model \citep{cordes02a} for the location of \mag\
by a factor of 5.  Accordingly, the remaining observations were made at
8.4\,GHz, where the effects of scatter-broadening are much less severe
and the astrometric precision much higher.


\begin{deluxetable}{cllcc}
\tabletypesize{\scriptsize}
\tablecaption{Summary of astrometric observations}
\tablewidth{0pt}
\tablehead{
\colhead{Observation date} & \colhead{Participating
telescopes\tablenotemark{a}} & \colhead{Observing bands\tablenotemark{b}}
& \colhead{Duration} & \colhead{Detection S/N} \\
\colhead{(yymmdd/MJD)} &  & \colhead{(MHz, dual polarization)} & \colhead{(hours)} &
}
\startdata
071111/54415 & AT, CD, HO, MP, PA     & 8409--8425,8425--8441                         &  5 &  6 \\
080327/54552 & AT, CD, HO, MP, PA     & 8409--8425,8425--8441,[8441--8457,8457--8473] & 12 & 11 \\
090226/54888 & AT, CD, HO, MP, PA     & 8409--8425,8425--8441,[8441--8457,8457--8473] & 10 & 20 \\
091211/55176 & AT, CD, HO, MP, PA, TI & 8409--8425,8425--8441,[8441--8457,8457--8473] &  8 &  9	
\enddata
\tablenotetext{a}{AT = Australia Telescope Compact Array, CD = Ceduna,
HO = Hobart, MP = Mopra, PA = Parkes, TI = Tidbinbilla.}
\tablenotetext{b}{Additional observing bands listed in square brackets were only
present at AT, MP, PA.}
\label{tab:obs}
\end{deluxetable}

Observing setups varied between epochs due to resource constraints
and equipment upgrades, with observation durations between 5 and 12
hours, and total bandwidths of 32\,MHz or 64\,MHz (dual polarization).
Table~\ref{tab:obs} summarizes the 8.4\,GHz observations, which were used
for the astrometric results detailed below.

At the commencement
of observations, the sparse calibrator grid in the southern skies meant
that no known phase reference sources were available within $6^{\circ}$
of \mag.  Instead, the observations were phase referenced to the bright
and compact radio source \cal\ ($\sim1$\,Jy at 8\,GHz).
\cal\ was identified as a probable calibrator based on
lower-resolution data and confirmed as such in the first astrometric epoch.
However, its relatively large angular separation from \mag, $5.\!^{\circ}3$,
means that the quality of phase referencing is much poorer than with other
equivalent LBA observations at 8.4\,GHz \citep[e.g.,][]{deller08b}.  A 5 minute
target/calibrator cycle was employed, with 2 minute calibrator scans
and 3 minute target scans.

After the first epoch, scans to check phase referencing were included on
other known calibrator sources within $6^{\circ}$ of \cal, and the results 
used to refine its position.  As the position for the calibrator source
\cal\ was refined in successive observations, different positions were
used in correlation at each epoch, but in the final analysis all
epochs were corrected to the more accurate position ultimately available
in the rfc\_2011d VLBI source catalog\footnote{http://astrogeo.org/vlbi/solutions/rfc\_2011d/},
(J2000.0) R.A.=$15^{\mathrm h}15^{\mathrm m}12.\!^{\mathrm s}67312$,
decl.=$-55\arcdeg59'32.\!^{\prime\prime}8361$,
with errors of 5.4\,mas in R.A. and 2.0\,mas in decl. 
This position was derived through observations in the LBA
Calibrator Survey, a dedicated southern hemisphere campaign
for absolute astrometry \citep{petrov11a}.

The data were correlated with the DiFX software correlator
\citep{deller07a}, using matched filtering (gating) on pulse profiles.
Pulsar ephemerides were supplied at each epoch from accompanying
timing observations done with the Parkes telescope, and were used to
set gates of widths that ranged between 7.5\% and 20\% of the pulse
period.  The applied gate width depended on the (varying) pulse profile at each epoch and the
uncertainty in the pulse phase prediction caused by unstable rotation.
Due to the relatively high observing frequency of 8.4\,GHz,
no ionospheric correction was applied after correlation.  During some
of the observations, the pulsar flux density varied significantly on
timescales of hours \citep[these variations are intrinsic to the magnetar;
see][]{crhr07,crj+08}.  The amplitudes were flattened (and the data
reweighted accordingly) using a ParselTongue script \citep{kettenis06a}
described in \citet{deller08a}.  Correction of the amplitudes in this
manner minimizes the flux scattered away from the true pulsar position and
maximizes the significance of the final detection, at the expense of an
elongated beam shape due to the time-variable weighting.  After imaging
using natural weighting, the AIPS task JMFIT was used to estimate
positions and errors.  The brightness of \mag\ varied considerably
between epochs, but it was detected in all epochs with a significance
ranging from 6\,$\sigma$ to 20\,$\sigma$ (Table~\ref{tab:obs}), leading
to positional fits with a nominal accuracy of $\approx 0.2-0.6$\,mas.

At 8.4\,GHz, the deconvolved, scatter-broadened angular diameter of \mag\
was found to be $\approx 6.5$\,mas.  In the four astrometric epochs,
fits to the semi-major axis ranged between 5.5 and 8\,mas.  This angular
broadening leaves very little flux on baselines with length 
$\ga1000$\,km, meaning these baselines contribute very little to the position
fit.  Consequently, the longest LBA baselines (to Hobart and Ceduna) 
contributed relatively little to the astrometric accuracy in this project.

\section{Astrometric results}
\label{sec:results}

Due to the large angular separation between the target and the calibrator,
the astrometric errors are dominated not by the signal-to-noise ratio but
by systematic contributions.  The predominant systematic effects are
introduced by the calibrator position uncertainty ($\sim5$\,mas), the
station position uncertainties (several centimeters) and the unmodeled
atmosphere (delay equivalent to a path length of several centimeters).
Both simulations \citep{pradel06a} and extrapolation from past results
with the LBA \citep{deller08b} suggest that under these conditions, the
systematic contribution to astrometric error should be of order 1\,mas.

For the astrometric fit, the magnetar parallax was fixed at 0.11\,mas,
corresponding to a distance of 9\,kpc \citep{crhr07}.  Varying the distance over a
wide range of values, from 2\,kpc to infinity, made an insignificant
($\sim0.1\,\sigma$) change to the proper motion fit. Without the
addition of any systematic error estimate, the reduced $\chi^2$ of the
fit was 8.0, indicating that as expected the formal errors in fitted position
considerably overestimate the astrometric accuracy.  The addition
of 1.2\,mas (R.A.) and 0.6\,mas (decl.) errors in quadrature was
sufficient to obtain a reduced $\chi^2$ of 1.0.  The optimal values
for the systematic error (minimizing the squared sum of the R.A.\ and
decl.\ components) were found via iterative minimization, following the
technique used by \citet{deller09b}.  The result of the final fit is shown
in Table~\ref{tab:fit}.  The absolute positional accuracy of the reference position for \mag\ is dominated
by the uncertainty in the absolute calibrator position.  The observing history
of the calibrator source is not long enough to determine if it exhibits
time-variable apparent position changes (which would corrupt the target
proper motion), but in this case the influence of such effects can be neglected,
since the magnitude of such changes are typically $< 0.1$\,mas\,yr$^{-1}$
\citep[e.g.,][]{macmillan07a}.

\begin{deluxetable}{lc}
\tabletypesize{\tiny}
\tablecaption{VLBI results for \mag}
\tablewidth{0pt}
\tablehead{
\colhead{Parameter} & \colhead{Fitted value and error} }
\startdata
R.A. (J2000)\tablenotemark{a}  & $15^{\mathrm h}50^{\mathrm m}
  54.\!^{\mathrm s}12386 \pm 0.\!^{\mathrm s}00005 \pm 0.\!^{\mathrm s}00064$       \\
Decl. (J2000)\tablenotemark{a} & $-54\arcdeg18'24.\!^{\prime\prime}1141
\pm 0.\!^{\prime\prime}0003 \pm 0.\!^{\prime\prime}0020$ \\
Epoch of position (MJD)                      & 54795.0                          \\
Proper motion in R.A., $\mu_{\alpha}\,{\rm cos}\,\delta$ & $ 4.8\pm0.5$\,mas\,yr$^{-1}$      \\
Proper motion in Decl., $\mu_{\delta}$ & $-7.9\pm0.3$\,mas\,yr$^{-1}$      \\
Total proper motion, $\mu$                & $ 9.2\pm0.6$\,mas\,yr$^{-1}$
\enddata
\tablecomments{All errors listed are $1\,\sigma$ confidence level intervals. }
\tablenotetext{a}{Celestial coordinates are given $\pm$ fit errors $\pm$
calibrator position errors. }
\label{tab:fit}
\end{deluxetable}

\begin{figure*}
\begin{center}
\begin{tabular}{cc}
\includegraphics[width=0.5\textwidth]{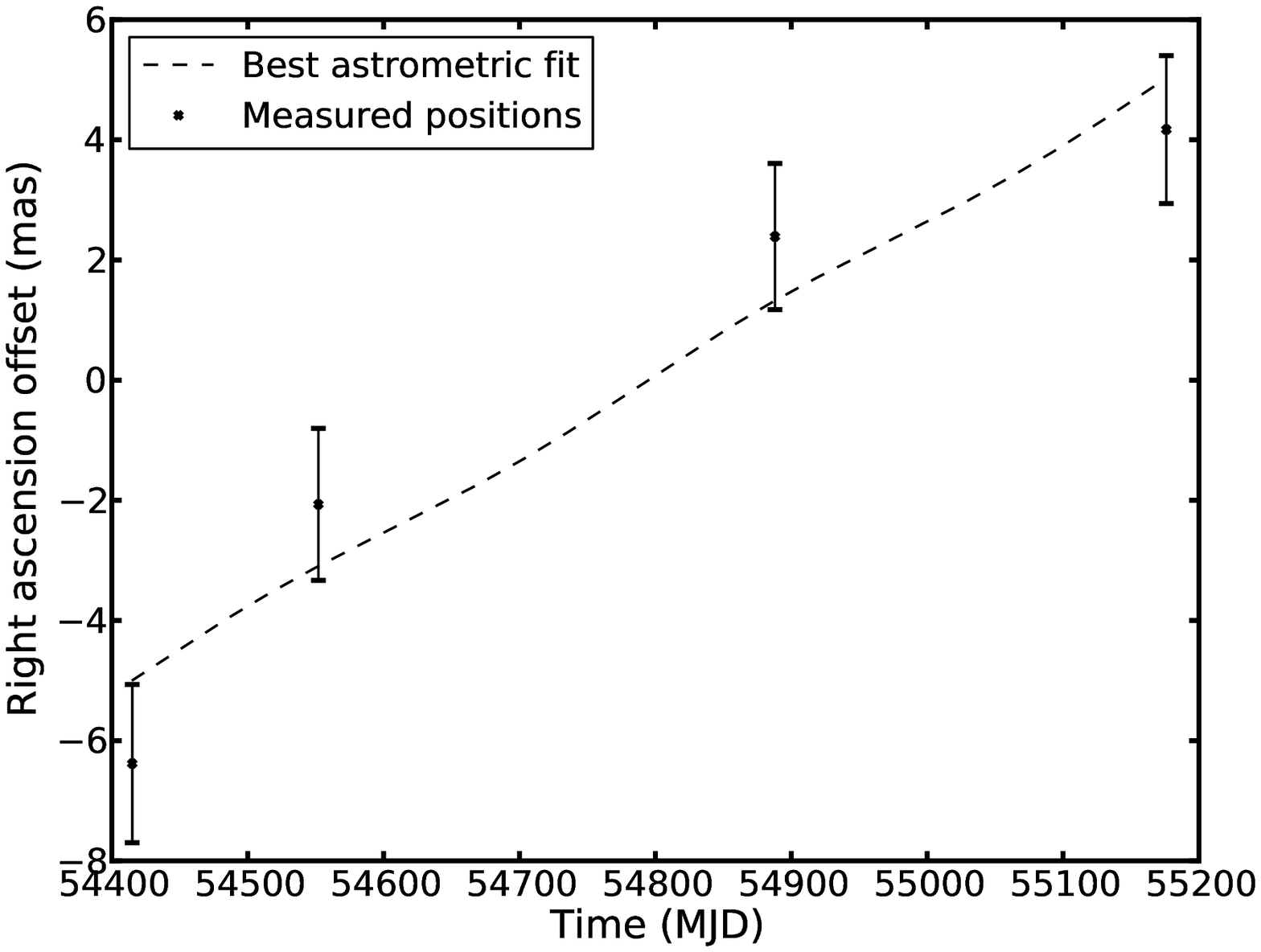}
\includegraphics[width=0.5\textwidth]{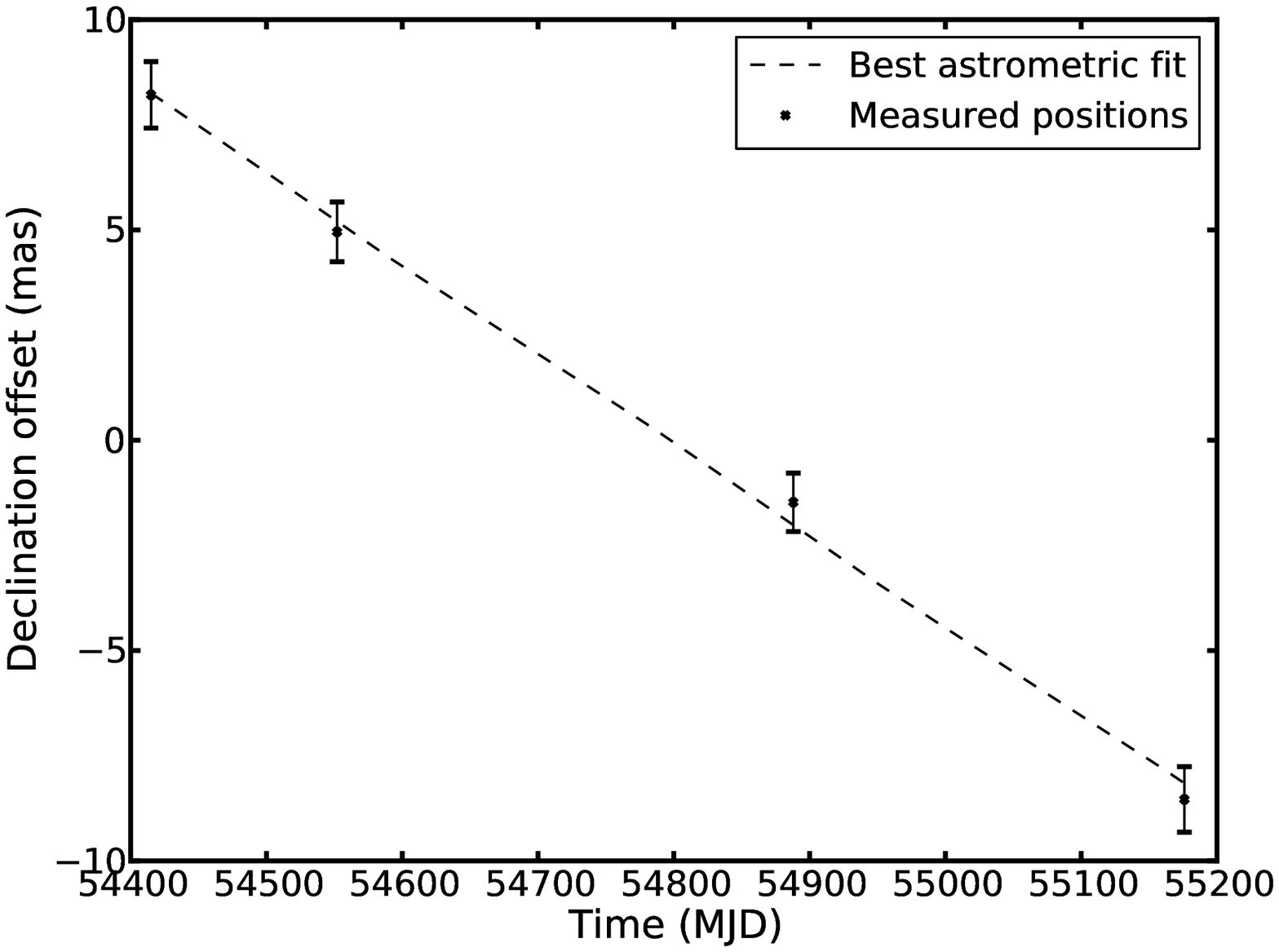}
\end{tabular}
\caption{Motion of \mag, with measured positions overlaid on the best
fit, and shown as offsets from the best-fit position at MJD~54795.0
(Table~\ref{tab:fit}).  {\bf (Left)} Motion in Right Ascension vs
time. {\bf (Right)} Motion in declination vs time.  }
\label{fig:fit}
\end{center}
\end{figure*}

Figure~\ref{fig:fit} shows the positional measurements and fitted motion
for \mag.  The accuracy is considerably better in decl.\ than in R.A.\
due to the predominantly north-south arrangement of antennas in the LBA.

\section{Velocity of \mag\ and of magnetars}
\label{sec:velocity}

Our results provide a secure measurement of the proper motion of \mag,
but converting this to a transverse velocity requires an estimate of
the magnetar's distance.  The free electron column density obtained from
the measured dispersion of the radio pulses, along with the electron
density model of \cite{cordes02a}, suggests that \mag\ is located at
$d=9$\,kpc \citep{crhr07}.  The uncertainty on this estimate is unknown,
but could approach 50\%.  An entirely different method relies on modeling
three dust-scattered X-ray rings observed following a huge outburst of the
magnetar that took place in 2009 January (one month before the third LBA
epoch listed in Table~\ref{tab:obs}).  Using the scattered X-rays,
\citet{tiengo10a} obtain a best-fit
distance of 4\,kpc, with anything in the range 4--8\,kpc plausible.

We therefore consider that a reasonable distance estimate for \mag\
is $d=6\pm2$\,kpc.  Along with our measured proper motion
of $\mu=9.2\pm0.6$\,mas\,yr$^{-1}$ (Table~\ref{tab:fit}), the
implied transverse velocity is $v_\perp = 260\pm90$\,km\,s$^{-1}$.
Correction for peculiar solar motion and Galactic rotation using a flat rotation curve and
the current IAU recommended rotation constants (R$_{0}$ = 8.5 kpc, $\Theta_{0}$ = 220\,km\,s$^{-1}$)
adjusts this slightly to 280$^{+130}_{-120}$\,km\,s$^{-1}$,
typical for the general population of ordinary pulsars
\citep[e.g.,][]{hobbs05a}.  This velocity is also comparable to
that obtained for the magnetar XTE~J1810--197 \citep[$v_\perp =
212\pm35$\,km\,s$^{-1}$ for $d=3.5\pm 0.5$\,kpc;][]{helfand07a}.
 The small angular velocity
of \mag\ also implies that an association between the pulsar and the
possible SNR shell G327.24--0.13 
that apparently surrounds it \citep{gg07}
remains plausible \citep[see][]{crhr07}. Since the characteristic
age of the pulsar is $\le 1400$~yr \citep{crj+08}, the pulsar could have moved
at most $\sim$13\arcsec\ since its birth.  This is consistent with its location
near the centre of the apparent SNR, which has a diameter of $4^{\prime}$.

With these two proper motions in hand, it is already very unlikely that
magnetars as a class have exceptionally high space velocities. For example,
if we assume that all magnetars have $v=1000$~km\,s$^{-1}$, then
the probability that one will be observed with $v_\perp\le 212$~km\,s$^{-1}$
is 0.023, while the probability that $v_\perp\le 280$~km\,s$^{-1}$
is 0.04.  Jointly, the probability of both observations is
$9 \times 10^{-4}$ under this assumption.  If all magnetars have
$v=500$~km\,s$^{-1}$, then the corresponding individual probabilities
become 0.094 and  0.171, with 0.016 for the joint probability.
We therefore consider $v<500$~km\,s$^{-1}$ as a reasonable
($\approx 2.2\sigma$) upper limit on the typical magnetar velocity.
Additional measurements or constraints on magnetar proper motions
would clearly be desirable to further constrain their distribution
of space velocities.

Unfortunately, X-ray observations with
{\it Chandra} have yielded only upper limits that are not very
restrictive.  \citet{kap09} find for 1E~2259+586
$\mu < 65\,{\rm mas\,yr}^{-1}, v_\perp < 930$\,km\,s$^{-1}$ assuming $d=3$\,kpc,
which is not as small a limit
as has already been inferred from the pulsar's central
location in its host SNR CTB 109.  For SGR 1900+14, \citet{kap09} find
$\mu < 54\,{\rm mas\,yr}^{-1}, v_\perp < 1300$\,km\,s$^{-1}$ assuming $d=5$\,kpc
(see also \citealt{del09}).  This result is not constraining if an association
of SGR 1900+14 with a young star cluster at $d=12.5$\,kpc is accepted \citep{vrb00,dav09},
since SGR 1900+14 remains within $15^{\prime\prime}$ of the center of the cluster.
Similarly, the apparent association of SGR 1806--20 with a cluster of giant
massive stars \citep{fuc99} requires that magnetar to have a small velocity.

The only other known radio-emitting magnetar, PSR~J1622--4950 \citep{lbb+10},
is an attractive target for the LBA.  But in the meantime, our LBA measurement of
the proper motion of \mag\ confirms that extremely high velocities need
not result as a consequence of the formation of magnetars, and suggests
that the velocity distribution of magnetars may not differ significantly
from that of ordinary, rotation-powered neutron stars.

\acknowledgements

The Long Baseline Array and the Parkes Observatory are part of the
Australia Telescope, which is funded by the Commonwealth of Australia
for operation as a National Facility managed by CSIRO.

\bibliographystyle{apj}

\end{document}